\documentclass[aps,prx,twocolumn,longbibliography,superscriptaddress,floatfix,nofootinbib]{revtex4-2}
\usepackage{epsfig,amsmath,amssymb,color,comment,physics}
\usepackage[makeroom]{cancel}
\usepackage[caption=false]{subfig}
\usepackage{mathrsfs}
\usepackage[countmax]{subfloat}
\usepackage[normalem]{ulem}
\usepackage[english]{babel}
\usepackage{dsfont}
\usepackage{float}
\usepackage[bookmarks=true,colorlinks,linkcolor=black,urlcolor=NavyBlue,citecolor=RoyalBlue]{hyperref}
\usepackage[dvipsnames]{xcolor}
\usepackage{mathtools}
\usepackage{siunitx}
\usepackage{upgreek}
\usepackage{comment}

\hypersetup{
    unicode=false,     
    pdftoolbar=false,  
    pdfmenubar=true,   
    pdffitwindow=false, 
    pdfstartview={FitH},
    pdftitle={},    
    pdfauthor={Authors},     
    pdfsubject={},   
    pdfcreator={},   
    pdfproducer={}, 
    pdfkeywords={quantum many-body scars} {superconducting processor} {quantum state tomography}, 
    pdfnewwindow=true,
    colorlinks=true,
    linkcolor=black,
    citecolor=blue, 
    filecolor=magenta,
    urlcolor=blue
}

\newcommand{\beginsupplement}{%
    \setcounter{table}{0}
    \renewcommand{\thetable}{S\arabic{table}}%
    \setcounter{figure}{0}
    \renewcommand{\thefigure}{S\arabic{figure}}%
    \setcounter{equation}{0}
    \renewcommand{\theequation}{S\arabic{equation}}%
    \setcounter{section}{0}
    \renewcommand{\thesection}{S\arabic{section}}%
   }

\newcommand{\Jnature}{Nature}

\newcommand{\Jpre}{Phys. Rev. E}

\newcommand{\Er}{E_{\textrm{r}}}

\newcommand{\mus}{\upmu\textrm{s}}

\begin{document}

\title{Observation of many-body dynamical localization}

\author{ Yanliang  Guo }
\thanks{These authors contributed equally to this work.}

\affiliation{Institut f{\"u}r Experimentalphysik und Zentrum f{\"u}r Quantenphysik, Universit{\"a}t Innsbruck, Technikerstra{\ss}e 25, Innsbruck, 6020, Austria}

\author{ Sudipta  Dhar}
\thanks{These authors contributed equally to this work.}

\affiliation{Institut f{\"u}r Experimentalphysik und Zentrum f{\"u}r Quantenphysik, Universit{\"a}t Innsbruck, Technikerstra{\ss}e 25, Innsbruck, 6020, Austria}

\author{ Ang Yang }
\thanks{These authors contributed equally to this work.}
\affiliation{School of Physics, Zhejiang University, Hangzhou 310027, China}

\author{ Zekai Chen}
\affiliation{Institut f{\"u}r Experimentalphysik und Zentrum f{\"u}r Quantenphysik, Universit{\"a}t Innsbruck, Technikerstra{\ss}e 25, Innsbruck, 6020, Austria}

\author{ Hepeng Yao}
\affiliation{DQMP, University of Geneva, 24 Quai Ernest-Ansermet, Geneva, CH-1211,  Switzerland}

\author{ Milena  Horvath}
\affiliation{Institut f{\"u}r Experimentalphysik und Zentrum f{\"u}r Quantenphysik, Universit{\"a}t Innsbruck, Technikerstra{\ss}e 25, Innsbruck, 6020, Austria}

\author{ Lei Ying}
\email{leiying@zju.edu.cn}
\affiliation{School of Physics, Zhejiang University, Hangzhou 310027, China}

\author{ Manuele  Landini}
\affiliation{Institut f{\"u}r Experimentalphysik und Zentrum f{\"u}r Quantenphysik, Universit{\"a}t Innsbruck, Technikerstra{\ss}e 25, Innsbruck, 6020, Austria}

\author{ Hanns-Christoph  N{\"a}gerl}\email{christoph.naegerl@uibk.ac.at}
\affiliation{Institut f{\"u}r Experimentalphysik und Zentrum f{\"u}r Quantenphysik, Universit{\"a}t Innsbruck, Technikerstra{\ss}e 25, Innsbruck, 6020, Austria}

\date{\today}

\begin{abstract}
The quantum kicked rotor is a paradigmatic model system in quantum physics. As a driven quantum system, it is used to study the transition from the classical to the quantum world and to elucidate the emergence of chaos and diffusion. In contrast to its classical counterpart, it features dynamical localization, specifically Anderson localization in momentum space. The interacting many-body kicked rotor is believed to break localization, as recent experiments suggest. Here, we present evidence for many-body dynamical localization for the Lieb-Liniger version of the many-body quantum kicked rotor. After some initial evolution, the momentum distribution of interacting quantum-degenerate bosonic atoms in one-dimensional geometry, kicked hundreds of times by means of a pulsed sinusoidal potential, stops spreading. We quantify the arrested evolution by analysing the energy and the information entropy of the system as the interaction strength is tuned. In the limiting cases of vanishing and strong interactions, the first-order correlation function exhibits a very different decay behavior. Our results shed light on the boundary between the classical, chaotic world and the realm of quantum physics.
\end{abstract}
\maketitle


Chaos is a phenomenon that is found almost everywhere in our daily life and it plays a central role in many of the sciences. Nonlinear complex systems, as they appear abundantly in mathematics, physics, biology, ecology, meteorology, economics, and other fields, are generally subject to chaotic dynamics~\cite{strogatz2018nonlinear}. The long-time evolution sensitively depends on the initial conditions and is inherently unpredictable. Chaos, although it has a negative connotation, is often very useful. It is intimately connected to ergodicity, and dense sampling of the available phase space drives many of the important physical processes, from thermodynamics to biology~\cite{baker1996chaotic}. However, this behavior is in stark contrast to what one expects from quantum physics~\cite{haake1991quantum,H-J1999}. The evolution of a closed (non-relativistic) quantum system is subject to the Schr\"odinger equation and it is unitary, i.e., in principle fully reversible and non-chaotic, at least to the point of measurement. The central questions are hence: Where is the boundary between the quantum coherent and the classical chaotic world? Which processes, apart from the measurement process, break the non-chaotic dynamics in a quantum system? Do particle-particle interactions necessarily lead to chaotic dynamics, in particular for strong interactions?


A minimal physical system to study chaotic behavior is a rotating object periodically kicked by an external force. A transition from integrable to chaotic motion is found as the kick strength crosses a critical value. Its quantum version, known as quantum kicked rotor (QKR), is a paradigmatic example in which quantum coherence can prevent the system from falling into the regime of quantum chaos even in the strong-driving regime, in stark contrast to the classical counterpart. This counter-intuitive phenomenon can be understood as Anderson localization in momentum space~\cite{anderson1958localization,fishman1982chaos,grempel1984quantum}. It is termed dynamical localization (DL) and has been demonstrated experimentally in dilute ultracold atomic gases~\cite{moore1995atom,manai2015experimental,Delande2009andersontransition}.

The QKR is a highly idealized single-particle model system. Realistic systems consist of many particles that interact with each other. Interactions almost always lead to a randomization of the many-body dynamics~\cite{srednicki1994,rigol2008}. Nevertheless, recent theoretical and experimental work has identified various dynamical quantum many-body systems for which ergodicity is broken. Examples include many-body localized cold atoms in lattices~\cite{Bloch-floquet-2017}, many-body scar states of interacting Rydberg atom arrays~\cite{QMBS2021}, and prethermalized samples of bosons in low dimension~\cite{Weiss-prethermal-2023, abanin2017rigorous}. Driving an interacting quantum many-body system further increases the complexity. For the case of the many-body QKR a breakdown of DL has been proposed using meanfield approaches \cite{Delande2020meanfield,shepelyansky1993delocalization,bodyfelt2011interactions}, and in fact recent cold-atom experiments with either weakly interacting one-dimensional (1D)~\cite{see2022many} or strongly interacting 3D gases \cite{cao2022interaction} align with these proposals. However, recent theoretical work \cite{rylands2020many,chicireanu2021dynamical,Vuatelet2021effectivethermalization,Vuatelet2023dynamicalmanybody,Fava2020} suggests that DL may persist in an interacting, even strongly interacting many-body quantum system. Specifically, interacting 1D bosonic gas as an ideal Lieb-Lininger QKR is proposed to support many-body dynamical localization (MBDL). Driving a strongly correlated interacting system surprisingly may not have to end up in chaotic dynamics.

\begin{figure}[ht!]
\raggedright
\includegraphics[width=\linewidth]{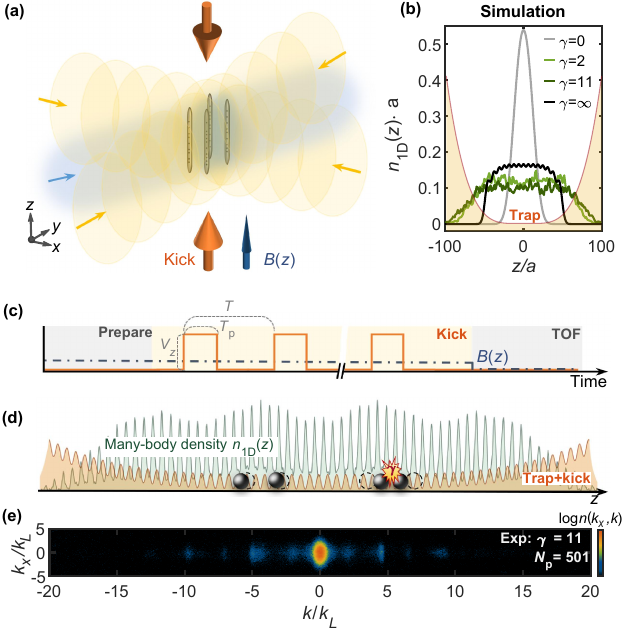}
\caption{
{\bf Experimental implementation of MBDL.}
(a) Sketch of the 3D BEC loaded into an array of 1D tubes generated by the 2D optical lattice (yellow) and partially compensated by a blue-detuned optical anti-trap (blue). The QKR lattice propagates along the vertical $z$-direction.
(b) Calculated ground-state many-body 1D spatial density $n_\mathrm{1D}(z)$ for the flat-bottom trap in units of $a^{-1}$ for various values of the interaction parameter $\gamma$. The data for $\gamma=2$ and $\gamma=11$ is obtained from a quantum Monte Carlo calculation at a temperature of $2$ nK, while the data for $\gamma=0$ and $\gamma=\infty$ are the zero-temperature results from a one-cycle Floquet map and fermionization method~\cite{rigol2005groundstate,wilson2020observation}.
(c) Schematic of the experimental QKR sequence with $N_\mathrm{p}$ pulses sandwiched between the preparation and TOF-imaging procedures.
(d) Illustration of $n_\mathrm{1D}(z)$ (green) for 4 interacting atoms in a cosine standing-wave potential in the flat-bottom trap (orange) for $\gamma\!=\!\infty$. The anti-nodes of $n_\mathrm{1D}(z)$ register with the nodes of the kick potential. The envelope of $n_\mathrm{1D}(z)$ with the 4 bumps reflects the presence of 4 atoms.
(e) Typical absorption image for $\gamma\!=\!11$ after $N_\mathrm{p}\!=\!501$ kicks showing the spatial density in color coding after TOF and integrated along the line-of-sight. }
\label{fig:system}
\end{figure}

Here we report the observation of MBDL in interaction-tunable 1D Bose gases. Using quantum-degenerate samples of Cs atoms, we realize the Lieb-Liniger quantum kicked rotor, a paradigmatic model for the simulation of driven interacting quantum many-body systems. We tune the interaction strength from the non-interacting regime to the Tonks-Girardeau (TG) regime~\cite{paredes2004tonks,kinoshita2004,haller2009} and find signatures of MBDL in a freezing of the momentum distribution and in a saturation of the energy and the entropy as the 1D Bose gases are kicked hundreds of times. The first-order correlation function exhibits a strikingly different decay in the two limiting cases of zero and strong interactions. Our results are in agreement with recent proposals
~\cite{rylands2020many,chicireanu2021dynamical,Vuatelet2021effectivethermalization}
and are captured qualitatively by our modeling.

The experiment starts by loading a Bose-Einstein condensate (BEC) of Cs atoms \cite{Kraemer2004} into an array of narrow 1D tubes created by a 2D optical lattice as illustrated in Fig.~\ref{fig:system}(a). The laser beams forming the lattice propagate in the horizontal $x$-$y$ plane at right angle to each other, and their power is adiabatically ramped up to give a lattice depth of $30 E_\mathrm{r}$, where $E_\mathrm{r}=\pi^2\hbar^2/(2ma^2)$ is the recoil energy. Here, $a\!=\!\lambda/2\!=\!\pi/k_\mathrm{L}\!=\!\!532.25$ nm is the lattice spacing with the wavenumber $k_\mathrm{L}$ that is set by the wavelength $\lambda$ of the lattice light. At this depth, tunneling between the tubes is fully suppressed, and the tubes can be considered as independent. A magnetic field $B$ along the vertical $z$-direction controls the inter-atomic interaction~\cite{weber2003} as quantified by $\gamma$, the dimensionless 1D Lieb-Liniger interaction parameter~\cite{cazalilla2011}. By choosing values between $\gamma=0$ and $\gamma=11$ we cover the range from the non-interacting, single-particle regime well into the strongly interacting TG regime in which the bosons have fermionized. The atoms are levitated against gravity by means of a gradient of $B$. The finite transversal extent of the lattice beams results in weak harmonic trapping with trap frequency $\nu_z\!=\!14.7(3)$ Hz along the tubes' direction. We partially compensate this trap by means of a horizontally propagating blue-detuned anti-trap laser beam. This forms a flat-bottom potential, as illustrated in Fig.~\ref{fig:system}(b). Typically, we fill about $10^4$ tubes, with a weighted average of $18$ atoms per tube. At this stage, the 1D systems are in equilibrium with a temperature around $2$ nK, which is estimated by the thermometry method demonstrated in Refs.~\cite{guo-cooling-2023,guo-crossoverD-2023}. Calculated 1D density distributions $n_\mathrm{1D}(z)$ for various values of $\gamma$ are included in Fig.~\ref{fig:system}(b).

\begin{figure*}
\centering
\includegraphics[width=\linewidth]{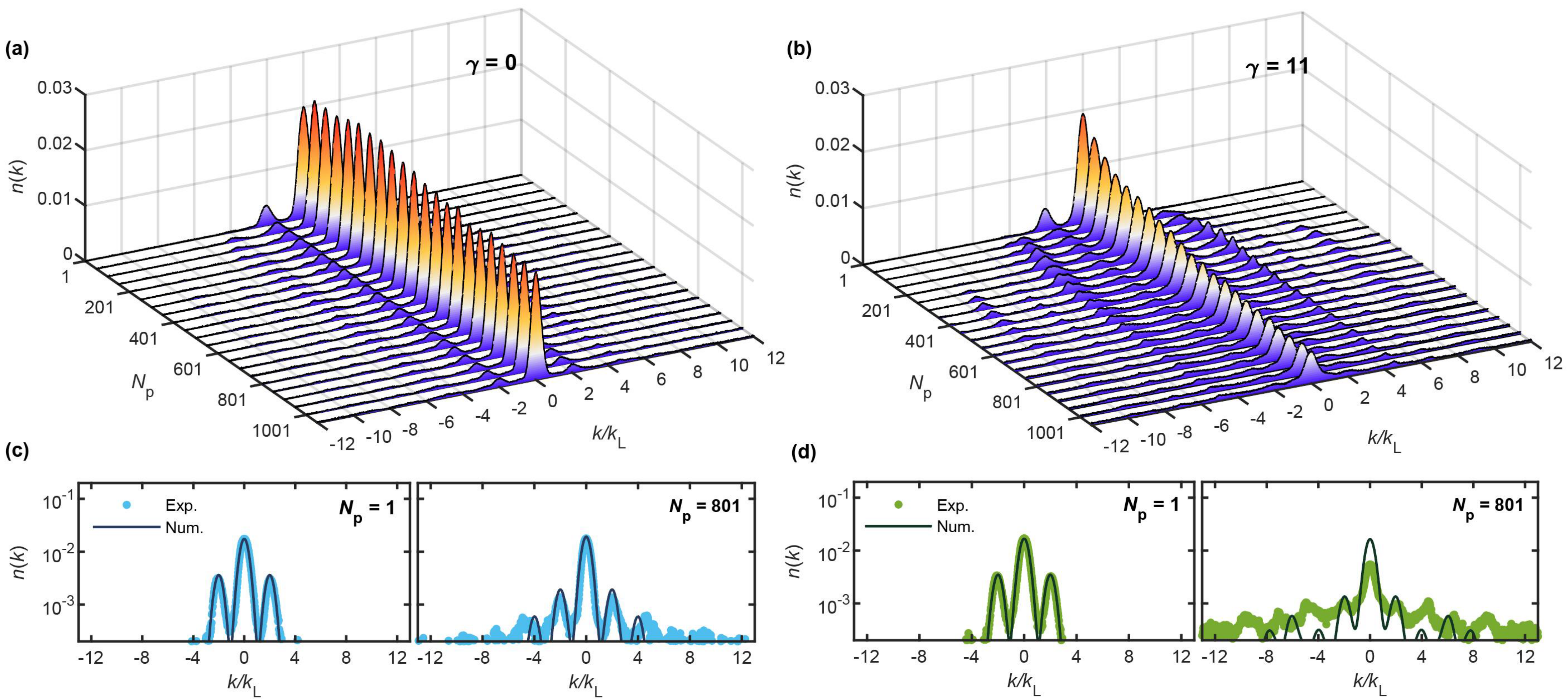}
\caption{{\bf Evolution of the 1D momentum distribution.} Measured $n(k)$ for the non-interacting (a) and strongly-interacting case (b) for a kick strength $K\!=\!3.3$. In (a) and (b) the number of kicks $N_\mathrm{p}$ is increased from $1$ to $1101$ in steps of $50$. Example distributions are shown in (c) and (d) for $N_\mathrm{p}\!=\!1$ and $801$ for $\gamma\!=\!0$ and $\gamma\!=\!11$, respectively. The data is compared to numerically calculated distributions for $\gamma\!=\!0$ and $\gamma\!=\!\infty$. In view of the finite TOF duration these are convoluted by Gaussians whose waist is half of the flat-bottom length. All experimental distributions are the average of three realizations.
}
\label{fig:momentum_evolution}
\end{figure*}

We next implement the QKR procedure. A longitudinal standing wave with lattice spacing $a$ and depth $V_z$ is periodically pulsed on with period $T$ and square-pulse duration $T_\mathrm{p}$ for a number of $N_\mathrm{p}$ pulses as illustrated in Fig.~\ref{fig:system}(c). We choose $T\!=\!60\,\mus$ and $80 \mus$, $T_\mathrm{p}\!=\!10\,\mus$, and $V_\mathrm{z}\!=\!20E_\mathrm{r}$, resulting in a dimensionless kick strength $K=3.3$ and $K=4.4$ respectively, see below and Supplemental Material. For values of $N_\mathrm{p}$ between $1$ and $1101$, the entire pulse sequence takes up to $66$ ms and $88$ ms, respectively. Fig.~\ref{fig:system} (d) illustrates the many-body wavefunction during the kicking procedure. We obtain the longitudinal momentum distribution $n(k)$ by a 20-ms time-of-flight (TOF) absorption-imaging technique. A typical absorption image is shown in Fig.~\ref{fig:system} (e) and $n(k)$ is obtained as a probability distribution by integration along the transversal direction and normalizing by the atom number. For a faithful measurement of $n(k)$ the particle interaction is switched off via a Feshbach-resonance's zero crossing \cite{weber2003} during TOF. From $n(k)$ we subsequently compute the kinetic energy $E$, the information entropy $S$, and the one-body correlation function $G$.

We model the system by the explicitly time-dependent Lieb-Liniger QKR Hamiltonian. For $N$ interacting bosons with mass $m$ moving in a 1D tube at zero temperature, it reads \cite{rylands2020many,Vuatelet2021effectivethermalization}
\begin{equation}
 \begin{split}
        H(t)=\sum_{i=1}^N\Bigg[  & \frac{P_i^2}{2m}+\hbar\kappa\cos{\left(2k_Lz_i\right)}\sum_{N_\mathrm{p}}\delta_{\tiny{T_\mathrm{p}}}\left(t-N_\mathrm{p}T \right) \\
        & + V_{\rm{ext}}\left(z_i\right)    \Bigg]
        + g_\mathrm{1D}\sum^N_{i<j}\delta(z_i-z_j).
 \end{split}
\label{original_H}
\end{equation}
Here, $\hbar\kappa\!=\!V_zT_\mathrm{p}/{2}$ sets the kick strength $K=8T\Er\kappa/\hbar$. The first three terms represent the QKR for a single particle at position $z_i$ moving in the trap potential $V_{\rm{ext}}$. The last summand models the $\delta$-type pairwise contact interaction whose strength $g_\mathrm{1D}$ is related to $\gamma$ via $\gamma=mg_\mathrm{1D}/n_\mathrm{1D}\hbar^2$. Details of our modeling are given in Supplemental Materials (SM).

We first turn to the evolution of the momentum distribution as $N_\mathrm{p}$ is increased. Fig.~\ref{fig:momentum_evolution} presents the experimental results of $n(k)$ for two limiting values of the interaction strength, namely for vanishing ($\gamma\!=\!0$) and for very strong interactions ($\gamma\!=\!11$). In the former case, the system is in the single-particle regime, in the latter it is deeply in the TG regime. Starting with the first kick, higher momentum states are populated, primarily the ones with $\pm 2 \hbar k_\mathrm{L}$. For the non-interacting case, some of the population is subsequently transferred to higher momentum states, but evidently further transfer is halted and the distribution becomes stationary, as can be seen from comparing the data sets with $N_\mathrm{p}\!=\!301$ and $N_\mathrm{p}\!=\!1001$. The system is dynamically localized \cite{manai2015experimental,Delande2009andersontransition,cao2022interaction}. In the interacting case, the change in the momentum distribution from $N_\mathrm{p}\!=\!1$ to about $N_\mathrm{p}\!=\!301$ is much more pronounced. However, the subsequent evolution, in particular from $N_\mathrm{p}\!=\!501$ to $N_\mathrm{p}\!=\!1101$, appears to have stopped. Examples of $n(k)$ after the first kick and in the localized state after $N_{\mathrm{p}}\!=\!801$ for the non-interacting and the strongly interacting cases are shown in Fig.~\ref{fig:momentum_evolution} (c) and (d), respectively. As we will argue below, the interacting system, after some initial evolution towards a broadened distribution, has entered the regime of MBDL.


\begin{figure*}
\centering
\includegraphics[width=\linewidth]{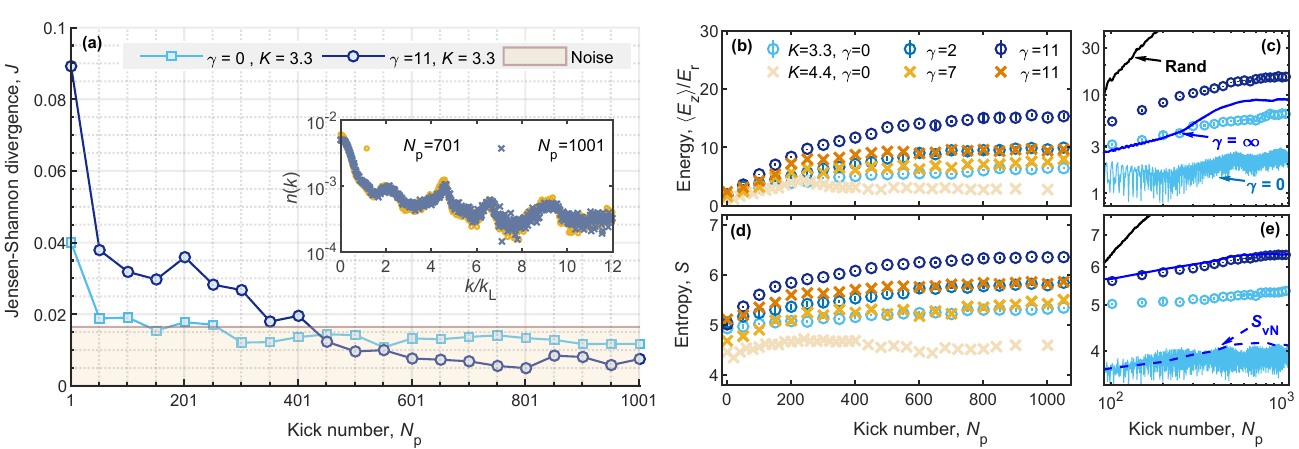}
\caption{{\bf Evidence for MBDL.} (a) Jensen-Shannon divergence $J$ between the momentum distributions at $N_\mathrm{p}$ and $N_\mathrm{p}+100$ as $N_\mathrm{p}$ is stepped in units of $50$, for the non-interacting ($\gamma\!=\!0$, light blue squares) and the strongly interacting ($\gamma\!=\!11$, dark blue circles) cases. The orange shaded area indicates the typical shot-to-shot noise determined by comparing 3 repeats at $N_\mathrm{p}\!=\!501$. Inset: Comparison of $n(k)$ at $N_\mathrm{p}=701$ and at $N_\mathrm{p}=1001$ for the strongly interacting case $\gamma\!=\!11$. (b-e) Time evolution of the kinetic energy $\langle E_z \rangle$ (b) and of the information entropy $S$ (d) for various values of the 1D interaction strength $\gamma$ as indicated and for two values of the kick strength $K\!=\!3.3$ ($T\!=\!60\,\mus$, circles) and $K\!=\!4.4$ ($T\!=\!80\,\mus$, crosses).
(c) and (e): Comparing the energy (c) and entropy (e) calculated for the non-interacting (light blue solid line) and infinitely interacting cases (dark blue solid line) to the data with $\gamma\!=\!0$ (light blue circles) and $\gamma\!=\!11$ (dark blue circles) for the case of $K=3.3$ on a log-log scale. Fast oscillations for the theory data for $\gamma\!=\!0$ are not resolved. For comparison, the time evolution of a non-interacting system with random kicks is presented (black line, average over 10 realizations). Additionally, the calculated von-Neumann entropy $S_{\mathrm{vN}}$ for infinite interactions is shown (dashed line). The statistical error in (b) through (e) is smaller than the symbol size.
}
\label{fig:energy_entropy}
\end{figure*}


We utilize the Jensen-Shannon divergence $J$ to quantitatively study the evolution of $n(k)$. It is a measure of the overlap of two probability distributions \cite{JSD1991} with values between $J\!=\!0$ (full overlap) and $J\!=\!1$ (no overlap), see SM. Fig.~\ref{fig:energy_entropy}(a) presents the time evolution of $J$ for both the non-interacting and the strongly interacting scenarios as $N_\mathrm{p}$ is stepped in units of 50. Initially, as the distributions still change, the value for $J$ is large compared to the noise floor. But quite quickly, $J$ drops towards the noise level. For the non-interacting case, the noise level is reached already after the first step, and overlap cannot be distinguished from noise beyond $N_\mathrm{p}\!=\!300$. The system has undergone dynamical localization. For the interacting case, this takes about $N_\mathrm{p}\!=\!450$ kicks. Afterwards, the change in the distributions is minimal and cannot be distinguished from the noise anymore. The system has undergone MBDL. This can also be seen from a direct comparison of two typical distributions taken at $N_\mathrm{p}\!=\!701$ and $N_\mathrm{p}\!=\!1001$ as shown in the inset of Fig.~\ref{fig:energy_entropy}(a). Within the error bars the two structured distributions are the same, and also the slight shifts away from the multiples of $2\hbar k_\mathrm{L}$ are well reproduced. This is remarkable, as the interacting system has been kicked 300 times more for the later measurement. Evidently, the interplay of quantum coherence and strong interactions prevents the breaking of localization.

We further quantify the behavior observed above by analyzing the longitudinal kinetic energy $\langle E_z \rangle \!\propto\! \langle k^2 \rangle$ and the information entropy $S\!=\!-\sum_{k}n(k)\mathrm{log}[n(k)]$ for various values of $\gamma$. The former is measure of the absorbed energy, and the latter relates to the degree of chaos in the system. The data is taken for two values of the kick strength, $K\!=\!3.3$ and $K\!=\!4.4$, and, next to the previous choice of $\gamma\!=\!0$ and $\gamma\!=\!11$, for intermediate values of $\gamma$. The experimental results are shown in Fig.~\ref{fig:energy_entropy}(b) and (d),  For any choice of $\gamma$, both observables, after an initial rise, settle to constant values after a few hundred kicks. Evidently, the evolution has stopped. The limiting values depend on $\gamma$: The larger $\gamma$, the larger the localization energy and the localization entropy for a given $K$. The limiting values also depend on $K$. One can see a clear difference for the data taken for $\gamma\!=\!11$, with a smaller kick strength giving larger localization values. We note that for a time evolution up to $N_\mathrm{p}\!=\!1001$, atom loss is small and transversal excitations are negligible. Stronger interactions lead to slightly higher atom loss, but losses remain below $25\%$. We also note that we can increase $\gamma$ to values that are much larger than the ones used here, e.g., to $\gamma\!=\!86$. For this choice we find increased atom loss and pronounced transversal excitations. These become evident already after about $100$ kicks. Details on loss and transversal excitations are discussed in SM.


In Fig.~\ref{fig:energy_entropy}(c) and (e) we compare the results for vanishing and very strong interactions to our theory data, obtained in the two limiting cases $\gamma\!=\!0$ and $\gamma\!\rightarrow\!\infty$, for which calculations can be carried out. While the experimental data lie above the theory predictions, the growth trend for the energy and the entropy for experiment and theory agree reasonably well. In particular, theory confirms the onset of localization after around $500$ kicks. The theory allows us to also calculate the von-Neumann entropy $S_\mathrm{vN}\!=\!-\mathrm{Tr}[\rho \mathrm{log}\rho]$, where $\rho$ is the one-particle density matrix (see SM) \cite{rigol2005groundstate,matthias2022ORDM}. It is a measure of the entanglement generated in the system. It also settles to a constant value, and its growth trend is almost the same as for the information entropy. All this behavior is in stark contrast to what happens when the system is subject to random kicks. In that case energy and entropy increase diffusively without bounds, analogous to the case of classical chaos.

We now turn to the quantum coherence of our kicked many-body system. It is contained in the one-particle density matrix $\rho(z,z')\!=\!\langle \hat{\Psi}^\dagger(z')\hat{\Psi}(z)\rangle$, where $\hat{\Psi}$ is the many-body field operator. It governs the functional form of the one-body correlation function via $G^{(1)}(z)\propto \int \rho(z,z') dz'$. In practice, one gets ${G}^{(1)}(z)$ from the Fourier transform of $n(k)$ \cite{paredes2004tonks,guo-crossoverD-2023}. The experimental results for ${G}^{(1)}(z)$ of the QKR are presented in Fig.~\ref{fig:strong_interaction&g1} for the TG regime and the non-interacting regime for two kick strengths and for consecutive values of $N_\mathrm{p}$ in the localized regime. As a consequence of the frozen momentum distribution, $G^{(1)}(z)$ does not depend on $N_\mathrm{p}$ after MBDL has occurred. The decay for the strongly interacting case is well fit by an exponential for intermediate distances. A frozen correlation length is considered to be a property of the effectively thermalized TG gas in MBDL phase, as predicted for trapping-free systems~\cite{Vuatelet2021effectivethermalization,Vuatelet2023dynamicalmanybody}. The exponential fit is not expected to cover $G^{(1)}$ for very short distances, which via  Fourier transform correspond to the region of large momenta. This is compatible with a predicted algebraic-like decay for large momenta in the MBDL phase~\cite{Vuatelet2023dynamicalmanybody,chicireanu2021dynamical,Vuatelet2021effectivethermalization}. Our data is in fact consistent with a $n(k)=\mathcal{C}_\mathrm{th} k^{-4}$ behavior. Here, the weight $\mathcal{C}_\mathrm{th}$ is Tan's contact of the effectively thermalized TG gas~\cite{tan2008b,Vuatelet2021effectivethermalization,Yao2018contact}, see details in SM. Contrary to the TG case, for the non-interacting system ${G}^{(1)}(z)$ exhibits a Lorentzian decay behavior. This implies an exponential decay in momentum space, signalling Anderson-like localization~\cite{anderson1958localization}. While the one-body correlation function for the MBDL phase requires more theoretical studies, especially for finite interactions and for trapped systems, the change in the behavior of the correlation function from one regime to the other highlights the many-body effect in the MBDL dynamics.

\begin{figure}
\centering
\includegraphics[width=\linewidth]{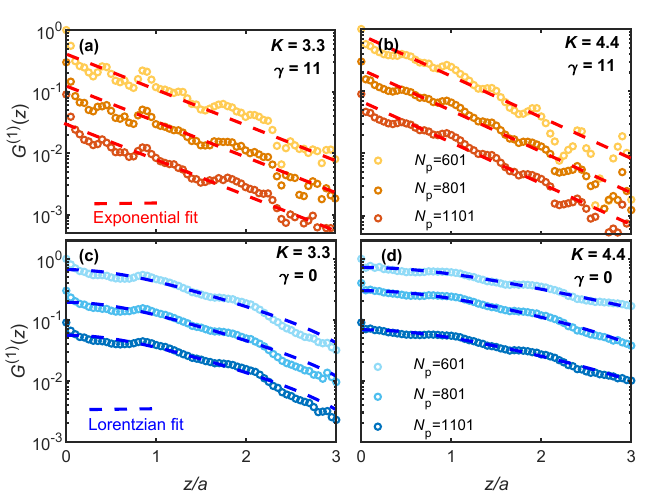}
\caption{
{\bf Decay of the one-body correlation function for DL and MBDL.}
(a) and (b): One-body correlation function $G^{(1)}(z)$ for strong interactions ($\gamma\!=\!11$) and two kick strengths and for $N_\mathrm{p}\!=\!601$, $801$, and $1101$, as indicated. The different datasets have been offset for better visibility. Exponential fits to the data for the range $z/a\!=\!0.25$ to $2.5$ are indicated by the dashed lines. (c) and (d): Same as above, but for the non-interacting case ($\gamma\!=\!0$). Lorentzian decays have been fit to the experimental data over the full range (dashed lines).
}
\label{fig:strong_interaction&g1}
\end{figure}

In summary, we have observed and analyzed a novel many-body phase termed MBDL by periodically kicking a TG gas confined in a flat-bottom trap. Our observation reveals a remarkable phenomenon: The momentum distribution of a strongly correlated system freezes despite the periodic kicks and preserves its characteristic interference structure. The kinetic energy and the information entropy show suppressed growth and saturation. Evidently, the quantum many-body system does not fall into the regime of chaos. The differences for the momentum distributions and the correlation functions between the non-interacting scenario and the strongly interacting one highlight the impact of the many-body effect under the periodic driving. Our findings raise a series of interesting questions to experiment and theory: What mechanisms could break MBDL in our Lieb-Liniger setting and how does chaos emerge from it?  What happens for intermediate interactions in the presence of inhomogeneous trapping for which integrability is expected to be broken? As interactions for our system are short-ranged in real space and hence necessarily long-ranged in momentum space, is there a mapping to real-space localized states for the case of long-range real-space interactions~\cite{Nandkishore2017,Defenu2023}? How is the relation to other many-body localized states that are, e.g., found in a lattice setting ~\cite{Adanin2019RMP} for which localization happens in real space? The observation of MBDL, in view of recent theoretical research~\cite{rylands2020many,chicireanu2021dynamical,Vuatelet2021effectivethermalization,Vuatelet2023dynamicalmanybody}, opens the door to a further exploration of the emergence of chaos in a many-body setting and provides new directions to investigate the boundary between the classical, chaotic world and the quantum realm, in particular for the strongly interacting scenario.

\noindent {\bf Acknowledgments}\\
The Innsbruck team acknowledges funding by a Wittgenstein prize grant under Austrian Science Fund (FWF) project number Z336-N36, by the European Research Council (ERC) under project number 789017, by an FFG infrastructure grant with project number FO999896041, and by the FWF's COE 1 and quantA. MH thanks the doctoral school ALM for hospitality, with funding from the FWF under the project number W1259-N27.
The team at Zhejiang University is supported by the National Natural Science Foundation of China (Grant No. 12375021) and the National Key R\&D Program of China (Grant No. 2022YFA1404203).
HY is also supported by the Swiss National Science Foundation under grant number 200020-188687. The numerical calculations make use of the ALPS scheduler library and statistical analysis tools~\cite{troyer1998,ALPS2007,ALPS2011}.

\noindent {\bf Data Availability}

\noindent The data that support the findings of this study are made publicly available from Zenodo by the authors at \cite{Zenodo}.





\bibliographystyle{apsrev4-2}

%

\cleardoublepage


\beginsupplement

\begin{center}
{\bf {\large Supplementary Materials}}\\
\end{center}

\begin{twocolumngrid}


\maketitle

\tableofcontents

\section{Experimental sequence}

The experiment starts with a 3D interaction-tunable Bose-Einstein condensate (BEC) consisting of approximately $1\!\times\!10^5$ Cs atoms prepared in the lowest magnetic hyperfine state $\vert F,m_F \rangle=\vert 3,3 \rangle$, held in a crossed-beam dipole trap and levitated against gravity by a gradient of the magnetic field $B(z)$~\cite{Kraemer2004}. The BEC is prepared in the Thomas-Fermi regime with a 3D s-wave scattering length of $a_{\textrm{\tiny 3D}}\!\approx\!190 a_0$, where $a_0$ is Bohr's radius. A 2D optical lattice is gradually ramped up within \SI{600}{\milli\second} to $30 E_\mathrm{r}$, cutting the 3D system in the horizontal $x$-$y$ plane into an array of 1D tubes oriented along the vertical $z$-direction. After loading the atoms into the lattice, the initial crossed-beam trap is ramped down within \SI{300}{\milli\second}. The offset magnetic field is then ramped up adiabatically to a desired value between $B\!\approx\!17$ G and $40$ G to set $a_{\textrm{\tiny 3D}}$ in the range between $a_{\textrm{\tiny 3D}}\!\approx\!0 a_0$ and $620 a_0$, with 3$a_0$ precision. After holding the atoms for an additional time of $100$~ms, the 1D tubes are periodically kicked by a pulsed one-dimensional optical lattice propagating along the longitudinal direction $z$. After the kicks, all trapping fields are switched off and the atoms perform a $20$-ms TOF, during which, importantly, the inter-particle interaction is turned off to avoid any residual interaction effects by switching $a_{\textrm{\tiny 3D}} $ to zero. This allows us to obtain the momentum distribution $n(k)$ as the various experimental parameters are varied via absorption imaging and integration over the transverse direction.

The horizontally propagating lattice beams cause weak harmonic trapping along the longitudinal direction of the tubes with a trap frequency of about $14.7$ Hz as a result of their finite beam size. We use an additional laser beam with a wavelength of $808$ nm, which is blue detuned to the laser-cooling transition and which propagates horizontally, to create an anti-trap in the $z$-direction. The waist of this beam is \SI{135}{\micro\meter}, whereas the lattice beams have a waist of \SI{300}{\micro\meter}. This results in a flat-bottom potential with a length of around \SI{50}{\micro\meter}. The detuning of the anti-trap beam is sufficiently large to not cause
any noticeable heating from spontaneous light scattering. \\

\section{Supplementary experimental data}

\subsection{MBDL with stronger interactions}
We have taken similar data as shown in Fig.~\ref{fig:energy_entropy} for an even stronger interaction strength $\gamma\!=\!86$. The experimental results are presented in Fig.~\ref{suppfig:larger_T}. For both two kick strengths, the qualitative evolution of the kinetic energy and of the entropy for $\gamma=86$ does not differ from the previous results in main text. However, the stronger interaction leads to more atom losses and to significant transversal excitations, as shown in Fig.~\ref{suppfig:atom_number}. Particularly for the case of $\gamma\!=\!86$ with $K\!=\!3.3$, the transversal waist increases substantially as a function of the kick number. These results indicate the situation that the gas enters into a regime of the dimensional crossover from 1D to 3D, which requires further exploration.


\begin{figure}[ht]
\raggedleft
\includegraphics[width=\linewidth]{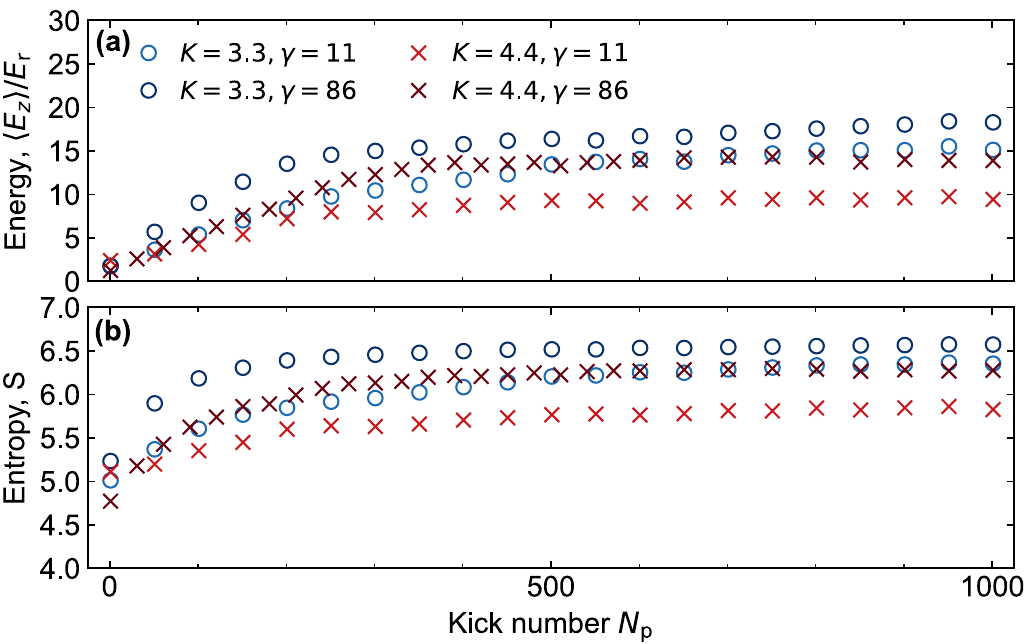}
\caption{Evolution of the kinetic energy (a) and of the information entropy (b) for $\gamma\!=\!11$ (same data as is shown in Fig.~\ref{fig:energy_entropy} of the main text) and $\gamma\!=\!86$ for two kick strengths, $K\!=\!3.3$ ($T\!=\!60\,\mus$) and $K\!=\!4.4$ ($T\!=\!80\,\mus$).
}
\label{suppfig:larger_T}
\end{figure}

\begin{figure}[ht]
\raggedright
\includegraphics[width=\linewidth]{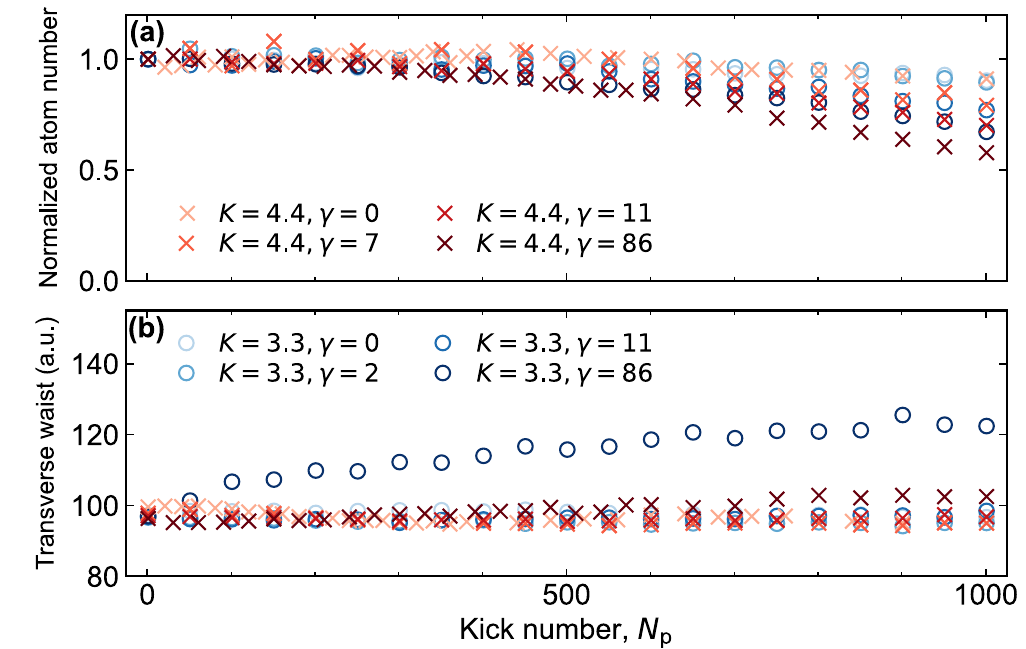}
\caption{
Evolution of the atom number and the transverse waist during the kicking process. (a) Atom number normalized to the initial one as a function of the kick number. (b) Evolution of the extracted Gaussian waist from the transverse momentum distribution of the main peak within $[-k_\mathrm{L}, k_\mathrm{L}]$. The data in (a) and (b) is shown for all cases from this work as indicated.
}
\label{suppfig:atom_number}
\end{figure}

\begin{figure}
\centering
\includegraphics[width=0.9\linewidth]{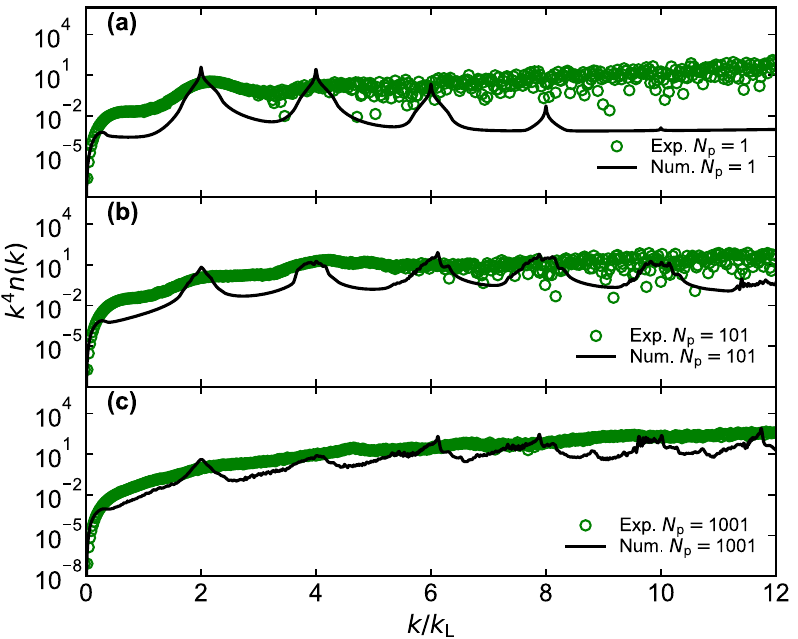}
\caption{
Distribution of $k^4n(k)$ for $\gamma\!=\!11$ (green circles) with $K\!=\!3.3$ at kick numbers $N_\mathrm{p}\!=\!1$ (a), $101$ (b), and $1001$ (c). The black solid lines show the corresponding numerical predictions for $\gamma\rightarrow\infty$.
}
\label{suppfig:tan's_contact}
\end{figure}

\subsection{Evidence of algebraic $k^{-4}$ tail during the kicking process}
In the MBDL phase, the momentum distribution of an untrapped kicked TG gas is expected to exhibit an algebraic-like decay at large momenta, namely $n(k)=\mathcal{C}_\mathrm{th} k^{-4}$, where the weight $\mathcal{C}_\mathrm{th}$ is expected to be Tan's contact of the effectively thermalized TG gas~\cite{Vuatelet2021effectivethermalization}. Figure~\ref{suppfig:tan's_contact} shows the distribution of $k^4n(k)$ of the kicked TG gas for different kick numbers for a probing of Tan's contact $\mathcal{C_\mathrm{th}}$. After $1$ kick, the quantity $k^4n(k)$ increases especially for $k\!>\!6k_\mathrm{L}$, whereas the numerical predictions show a clear plateau (see also Fig.~\ref{suppfig:Num_G1&nk}(c)). When $N_\mathrm{p}\!=\!101$, at a point in time at which the energy and entropy are still increasing, we find that the measured $k^4n(k)$ fluctuates around a constant at large momenta. As the kick number increases into the localized regime, the measured $k^4n(k)$ presents a growth trend at small momenta while it nearly approaches a plateau for momenta beyond $6k_\mathrm{L}$, in agreement with the corresponding numerical predictions. This possibly indicates that the momentum distribution in the MBDL phase exhibits a power-law decay $\mathcal{C_\mathrm{th}}k^{-4}$ at large momenta. This requires further investigations.


\subsection{Many-body QKR model and fermionization of bosons}
Our numerical calculations are based on the Hamiltonian of Eq.~(\ref{original_H}) in the main text. We introduce a set of dimensionless parameters to rewrite the Hamiltonian~\cite{Delande2009andersontransition}:
\begin{equation}
    \begin{split}
        \zeta &= 2k_\mathrm{L}z, \ \ \ \
        p = \frac{2k_\mathrm{L}TP}{m}, \ \
        \tau  = \frac{t}{T},  \\
        \hbar_{\rm{eff}} &= \frac{8TE_\mathrm{r}}{\hbar}, \ \
        K = \hbar_{\rm{eff}}\kappa, \ \
        g = \frac{8k^3_\mathrm{L}T^2g_\mathrm{1D}}{m},\\
    \end{split}
\label{scaled_parameter}
\end{equation}
where $E_\mathrm{r}={\hbar^2 k_\mathrm{L}^2}/{2m}$ is the recoil energy and $\hbar_{\rm{eff}}$ denotes the effective Planck constant. This scaling makes $p=\hbar_\mathrm{eff}$ corresponding to $P=2\hbar k_\mathrm{L}$, which allows us to take $k_\mathrm{L}$ as the unit of momentum. Then, the dimensionless Hamiltonian with $\delta$-function kicks is written as
\begin{equation}
    \begin{aligned}
        \mathcal{H}(\tau) &= \sum_i^N\left(\frac{p_i^2}{2} + \mathcal{V}_{\rm{ext}}(\zeta_i) + K\cos{(\zeta_i)}\sum_{N_\mathrm{p}}\delta(\tau-N_\mathrm{p}) \right)\\
        &+ g\sum^N_{i<j}\delta(\zeta_i-\zeta_j).
    \end{aligned}
\label{dimensionless_H}
\end{equation}
As $g$ is proportional to the Lieb-Liniger parameter $\gamma$, we use $\gamma$ to quantify the interaction strength throughout the paper.

For the non-interacting scenario, the Hamiltonian is simplified to $ \mathcal{H}_{\rm{QKR}}(\tau)= \mathcal{H}(\tau,\gamma=0)$.
After evolving the ground state $|\psi\rangle$ of bosons in the external trap potential $\mathcal{V}_{\rm{ext}}$, we
use a split-step fast Fourier transform method to alternate between the real-space basis and the momentum-space basis. Here, space is discretized before the stroboscopic evolution and the Floquet operator over one period is written as
\begin{equation}
    {U}=\exp{-i\left(\frac{\hbar_{\rm{eff}}k^2}{2}+\frac{\mathcal{V}_{\rm{ext}}(\zeta)}{\hbar_{\rm{eff}}}\right)}
    \exp{-i\frac{K}{\hbar_{\rm{eff}}}\cos{(\zeta)}},
\label{floquet_map}
\end{equation}
which generates the stroboscopic evolution $|\psi(\tau+1)\rangle=U|\psi(\tau)\rangle$.
Applying $U$ repeatedly, we obtain the long-time dynamic observables of the ensemble of bosons, in particular, the momentum distribution $n(k,\tau)$ and the averaged kinetic energy $\langle k^2(\tau)\rangle$.

The kick above is an ideal delta pulse, denoted by $\delta(t-N_\mathrm{p}T)$. In our experiments and numerics, we apply a square pulse with finite pulse duration $T_\mathrm{p}$:
\begin{equation}
    \delta_{T_\mathrm{p}}(t-N_\mathrm{p}T) =
    \begin{cases}
  \frac{1}{T_\mathrm{p}}  & |t-N_\mathrm{p}T|\leq T_\mathrm{p}, \\
  \ 0 &  |t-N_\mathrm{p}T| > T_\mathrm{p}.
\end{cases}
\end{equation}
Thus, the dimensionless Floquet operator over one period can be written as
\begin{equation}
\begin{split}
    U=&\exp{-i\left(\frac{\hbar_{\rm{eff}}k^2}{2}+\frac{\mathcal{V}_{\rm{ext}}(\zeta)}{\hbar_{\rm{eff}}}\right)\frac{T-T_\mathrm{p}}{T}} \\
    \times&\exp{-i\frac{K}{\hbar_{\rm{eff}}}\cos{(\zeta)}-i\left(\frac{\hbar_{\rm{eff}}k^2}{2}+\frac{\mathcal{V}_{\rm{ext}}(\zeta)}{\hbar_{\rm{eff}}}\right)\frac{T_\mathrm{p}}{T}}.
\end{split}
\label{modify_floquet_map}
\end{equation}



For the scenario of strong interactions in the TG regime ($\gamma \to \infty$), the strong local repulsion leads to the situation that only one particle can occupy one point in space $z$. Utilizing the Bose-Fermi mapping, we can straightforwardly compute the exact time-dependent many-body wavefunction of bosons $\Psi_B(\{\zeta\},\tau)$ at positions $\{\zeta\}=\{\zeta_1,\zeta_2,\dots,\zeta_N\}$~\cite{buljan2008bose-fermimap,Vuatelet2023dynamicalmanybody}
\begin{equation}
    \Psi_B(\{\zeta\},\tau) = \prod \limits_{i<j} \mathrm{sgn}(\zeta_i-\zeta_j) \Psi_F(\{\zeta\},\tau),
\label{bosonic_wave_function}
\end{equation}
where $\Psi_F(\{\zeta\},\tau)=\rm{det}[\psi_i(\it{\zeta}_j,\tau)]/{\sqrt{N!}}$ is the many-body wavefunction of free fermions expressed in terms of the Slater determinant of single-particle wavefunctions $\psi_i(\zeta_j,\tau)$.

We then calculate the one-particle density matrix ~\cite{Vuatelet2021effectivethermalization,Vuatelet2023dynamicalmanybody},
\begin{equation}
    \begin{aligned}
        \rho(\zeta,\zeta^{\prime},\tau) = N\int d\zeta_2\dots d\zeta_N &\Psi_B(\zeta,\zeta_2\dots \zeta_N,\tau)\times\\
                     &\Psi_B(\zeta^{\prime},\zeta_2\dots \zeta_N,\tau),
    \end{aligned}
\label{density_matrix}
\end{equation}
to obtain the density and momentum distributions of the fermionized Bose gas. The diagonal part of the matrix is the spatial density and the momentum distribution is given by
\begin{equation}
    n(k,\tau) = \int d\zeta d\zeta^{\prime}e^{ik(\zeta-\zeta^{\prime})}\rho(\zeta,\zeta^{\prime},\tau).
\label{momentum_distribution}
\end{equation}
Furthermore, the correlation function $G^{(1)}(z)$ is defined as average of $\rho(\zeta^{\prime},\zeta^{\prime\prime})$ over all distances~\cite{Xu2015universal,Vuatelet2021effectivethermalization}, and it is written as
\begin{equation}
\begin{split}
    G^{(1)}(\zeta)
    =\frac{1}{L_z}
    \bigg[
     & \int_0^{L_z}\rho \left(  \zeta^{\prime},\zeta^{\prime}+\zeta              \right)  d\zeta^{\prime}  \\
   + & \int_0^{L_z}\rho \left(  \zeta^{\prime\prime}+\zeta,\zeta^{\prime\prime}  \right)  d\zeta^{\prime\prime}
    \bigg],
\end{split}
\end{equation}
which can be obtained from the Fourier transform of the momentum distribution~\cite{paredes2004tonks,guo-crossoverD-2023}.

In practice, we employ the 1D hard-core Bose-Hubbard model to exactly compute the one-particle Green's function $G_{ij}=\langle \hat{b}_i\hat{b}_j^{\dag}\rangle$ by using the Jordan-Wigner transformation (JWT)~\cite{rigol2005groundstate,buljan2008bose-fermimap}
\begin{equation}
    \hat{b}_i^{\dag} = \hat{f}_i^{\dag}\prod \limits_{\beta-1}^{i-1}e^{-i\pi\hat{f}_{\beta}^{\dag}\hat{f}_{\beta}},\quad \hat{b}_i = \prod \limits_{\beta-1}^{i-1}e^{i\pi\hat{f}_{\beta}^{\dag}\hat{f}_{\beta}}\hat{f}_i,
\label{Jordan-Wigner_transform}
\end{equation}
where ${f}_i^{\dag}$ and ${f}_i$ are the creation and annihilation operators for spinless fermions. The bosonic creation $\hat{b}^{\dag}$ and annihilation $\hat{b}$ operators satisfy the on-site anti-commutation rule $\{\hat{b}_i,\hat{b}_i^{\dag}\}=1$, while they commute on different sites $[\hat{b}_i,\hat{b}_j^{\dag}]=0$. In the end, the one-particle density matrix correspondingly writes
\begin{equation}
    \rho_{ij} = \langle \hat{b}_i^{\dag}\hat{b}_j\rangle = G_{ij}+\delta_{ij}(1-2G_{ii}).
\label{discreted_density_mat}
\end{equation}
\\

\section{Numerical methods}
\subsection{Details of exact numerics}
We model the experimental system as a one-dimensional tube with a fixed number of atoms, described by the periodic many-body Hamiltonian in the presence of a confining potential $\mathcal{V}_\mathrm{ext}$ as presented in the main text. In our experiment, the weighted averaged number of atoms per tube can be estimated by $\overline{N}=\sum_{x_i,y_i}N^2_{x_iy_i}/\sum_{x_i,y_i}N_{x_iy_i}\!\approx\!18$, so we always choose $N\!=\!18$ unless stated otherwise. The dimensionless potential $\mathcal{V}_\mathrm{ext}$ takes the form
\begin{equation}
    \mathcal{V}_\mathrm{ext}(\zeta) = \frac{4k^2_\mathrm{L}T^2}{m}\left[V_1\left(1-e^{-\frac{\zeta^2}{2k^2_\mathrm{L}w^2_1}}\right) + V_2\left(e^{-\frac{\zeta^2}{2k^2_\mathrm{L}w^2_2}} - 1\right)\right],
\end{equation}
which is a sum of a Gaussian trap and a Gaussian anti-trap. The parameters are $V_1\approx45.7E_\mathrm{r}$, $V_2\approx9.3E_\mathrm{r}$, $w_1\!=$ \SI{300}{\micro\meter} and $w_2\!=$ \SI{135}{\micro\meter}.

Here we focus on the calculations in the TG limit ($\gamma\rightarrow\infty$). The Bose-Fermi mapping allows us to compute the exact many-body wave function of bosons and calculate the corresponding correlation functions in trapped systems both in equilibrium and far from equilibrium~\cite{wilson2020observation}. This mapping directly leads to many similar quantities between strongly interacting bosons and free fermions, for example, the kinetic energy and the entropy~\cite{paredes2004tonks}. However, there are still several non-local observables that are strongly modified due to the symmetry of the bosonic wavefunction, e.g. the one-particle density matrix, since fermionized bosons do not obey the Pauli exclusion principle in momentum space.

In detail, the simulation involves following steps for two the limiting scenarios:

For the non-interacting limit ($\gamma\!=\!0$), the number of atoms is irrelevant and all bosons initially occupy the ground state.

(i) We discretize space to obtain the matrix form of the single-particle Hamiltonian with dimension $N_\mathrm{dim}\times N_\mathrm{dim}$, and then carry out exact diagonalization in the real-space basis to obtain the ground state $|\psi_0\rangle$ of the system.

(ii) Starting from the ground state, we evolve it by the Floquet operator using a split-step fast Fourier transform method to alternate between the real-space basis and the momentum-space basis. This allow us to obtain the momentum distribution $n(k,\tau)={\lvert\langle k|\psi_0(\tau)\rangle\rvert}^2$ and kinetic energy $E_z(\tau)=\langle\psi_0(\tau)|k^2|\psi_0(\tau)\rangle$.

For the TG limit ($\gamma\rightarrow\infty$), the number of atoms is fixed at $N\!=\!18$. The $N$ bosons are considered being fermionized, meaning that they cannot occupy the same eigenstate because of the Pauli exclusion principle.

(i) We exactly diagonalize the single-particle Hamiltonian in the real-space basis to obtain the lowest $N$ eigenstates (i.e. $\{|\psi_i\rangle\}, i\!=\!0, 1, \cdots, N-1$) of the system.

(ii) These $N$ eigenstates are evolved separately according to the single-particle Hamiltonian, giving us $N$ time-dependent wavefunctions $|\psi_i(\tau)\rangle$ with $i=0,1,\cdots,N-1$. The dynamics of the kinetic energy $E_z(\tau)=\sum_i\langle\psi_i(\tau)|k^2|\psi_i(\tau)\rangle$ for the TG gas is simply the sum of the kinetic energies of these $N$ fermions.

(iii) Since we have the $N$ wavefunctions of the free fermions, we calculate the one-particle Green's function $G_{ij}(\tau)=\langle \Psi_B(\tau)| \hat{b}_i\hat{b}_j^{\dag}|\Psi_B(\tau)\rangle$ for the bosons. Through the Jordan-Wigner transformation~\cite{rigol2005groundstate,buljan2008bose-fermimap}, we have
\begin{equation}
    G_{ij}(\tau) = \left\langle \Psi_{F}(\tau)\right|\prod_{\beta-1}^{i-1}e^{i\pi\hat{f}_{\beta}^{\dag}\hat{f}_{\beta}}\hat{f}_i \hat{f}_j^{\dag}\prod_{\mu-1}^{j-1}e^{-i\pi\hat{f}_{\mu}^{\dag}\hat{f}_{\mu}}\left|\Psi_{F}(\tau)\right\rangle,
\end{equation}
where $\vert\Psi_{F}(\tau)\rangle=\prod_i^N\sum_\sigma^{N_\mathrm{dim}}\psi_{i\sigma}(\tau) \hat{f}_{\sigma}^{\dag}\vert 0\rangle$ is the fermionic many-body wavefunction and $\psi_{i\sigma}(\tau)$ corresponds to the $\sigma$-th element of the wavefunction $\psi_i(\tau)$~\cite{rigol2005groundstate}. Furthermore, we obtain the one-particle density matrix $\rho_{ij} = \langle \hat{b}_i^{\dag}\hat{b}_j\rangle = G_{ij}+\delta_{ij}(1-2G_{ii})$ and the other related observables such as the density (diagonal part), the momentum distributions (Fourier transform) and the von-Neumann entropy $S_\mathrm{vN}=-\mathrm{Tr}\rho \mathrm{log}\rho$. The correlation function $G^{(1)}$ is then determined by the Fourier transform of the momentum distributions.

We typically use a basis size of $N_\mathrm{dim}\!=\!10000$ and a longitudinal length $L_\mathrm{z}\!=\!$ \SI{300}{\micro\meter}. As the lattice is at very low fillings ($N/N_\mathrm{dim}\rightarrow0$), the system is effectively in the continuum~\cite{wilson2020observation}.

\subsection{Quantum Monte Carlo simulation}

We use the quantum Monte Carlo (QMC) method to simulate the equilibrium state of the 1D bosons, for computing its density profile and interaction strength $\gamma \sim g/n_\mathrm{1D}$. The method is path integral Monte Carlo~\cite{ceperley-PIMC-1995} with worm-algorithm implementations~\cite{boninsegni-worm-short-2006,boninsegni-worm-long-2006}, similar to the implementation in Refs.~\cite{guo-crossoverD-2023,guo-cooling-2023}. Here, we simulate the equilibrium state of one weighted average tube in the presence of the continuous flat-bottom potential within the grand-canonical ensemble. We firstly estimate the temperature of the system in equilibrium based on the thermometer proposed in ~\cite{guo-crossoverD-2023}, according to the data of the correlation function $G^{(1)}$. Then, with the estimated temperature, we compute the density profile directly from the configuration averaging in the real space. The results are shown in Fig.~1(b).\\

\subsection{Calculation of Jensen-Shannon divergence}
The Jensen-Shannon divergence $J$ is commonly used to determine the similarity of two probability distribution \cite{JSD1991}. In the main text, we use this statistical tool to quantify to what extend the momentum distribution evolves. For two probability distributions $P(k)$ and $Q(k)$ it is given by
\begin{equation}
    \begin{split}
    J= \frac{1}{2}\sum_k   \bigg[ & P(k)\log_2 \frac{2P(k)}{P(k)+Q(k)}  \\
    & +  Q(k)\log_2\frac{2Q(k)}{P(k)+Q(k)}  \bigg].
    \end{split}
\end{equation}
The value of $J$ is between 0 ($P$ and $Q$ have full overlap) and 1 ($P$ and $Q$ have no overlap).


\section{Additonal numerical results}

\begin{figure}
\raggedright
\includegraphics[width=\linewidth]{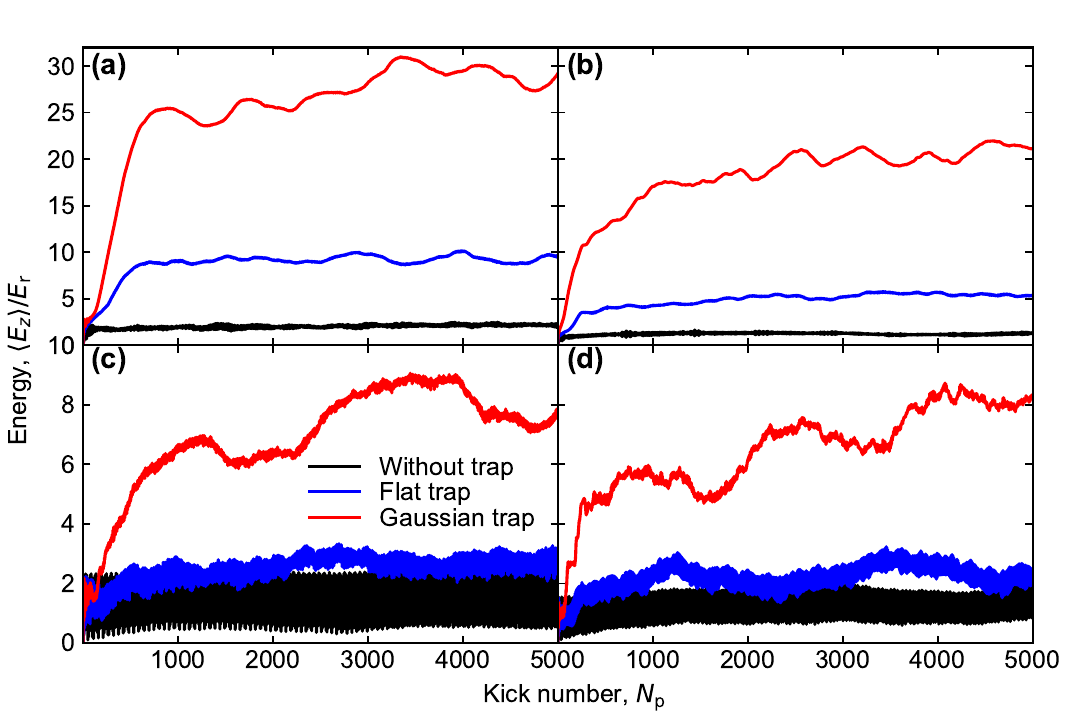}
\caption{
Evolution of the calculated kinetic energy per particle with $K\!=\!3.3$ for $\gamma\rightarrow\infty$ (a) and $\gamma\!=\!0$ (c) for various trapping conditions as indicated. (b) and (d) are the same as (a) and (c) but with $K\!=\!4.4$.
}
\label{suppfig:Num_trap}
\end{figure}

\begin{figure}
\centering
\includegraphics[width=0.9\linewidth]{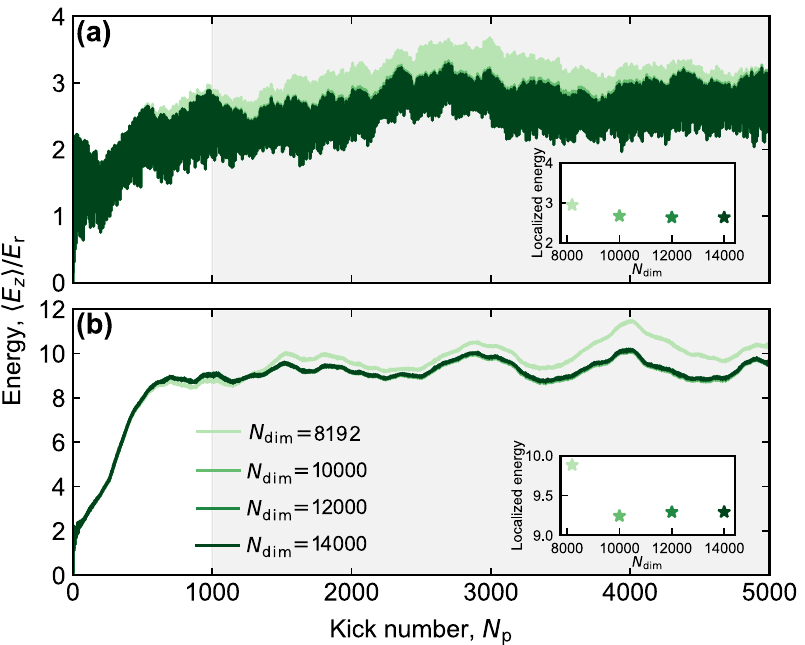}
\caption{
Evolution of the calculated kinetic energy per particle with $K\!=\!3.3$ for $\gamma\!=\!0$ (a) and $\gamma\rightarrow\infty$ (b) for different sizes of the basis set ranging from $2^{13}$ to 14000. Insets in (a) and (b): localized energy averaged from the grey region in (a) and (b) as a function of the basis size $N_\mathrm{dim}$.
}
\label{suppfig:Num_dimension}
\end{figure}

\begin{figure}
\centering
\includegraphics[width=\linewidth]{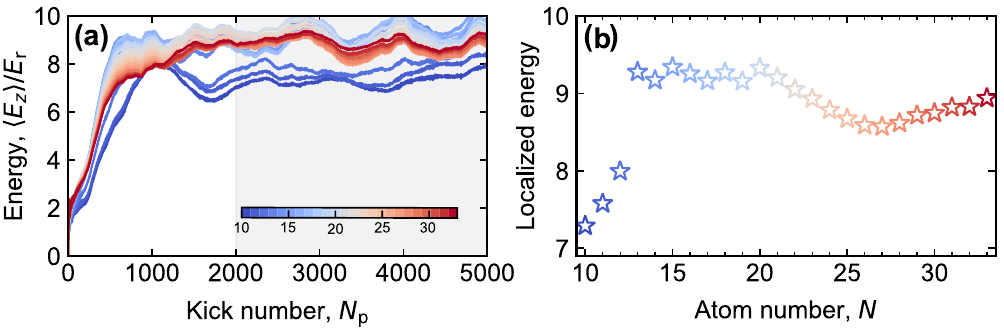}
\caption{
(a) Calculated kinetic energy per particle evolution for $K\!=\!3.3$ and $\gamma\rightarrow\infty$ for different atom numbers in a range from 10 to 33. (b) Localized energy averaged from the grey region in (a) as a function of atom number $N$.
}
\label{suppfig:Num_particle_number}
\end{figure}

\begin{figure}
\raggedright
\includegraphics[width=\linewidth]{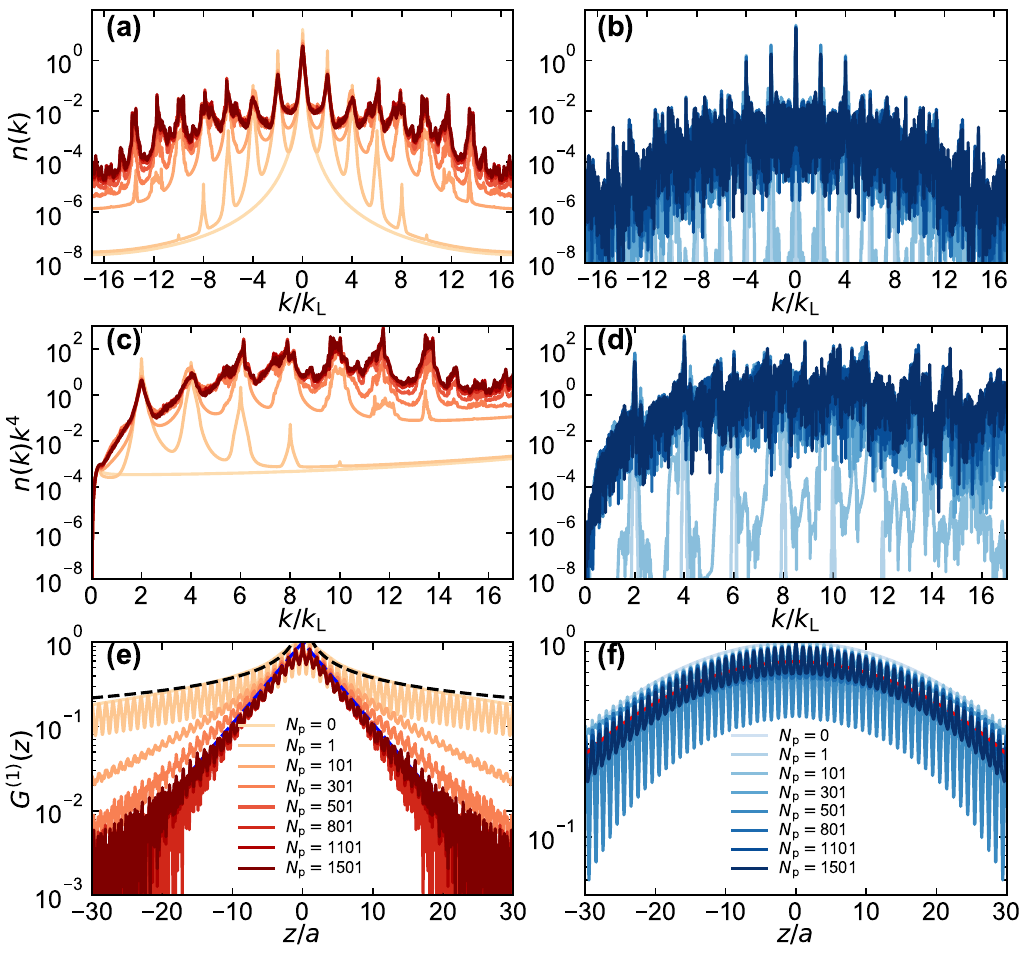}
\caption{
Calculated distributions $n(k)$ and $n(k)k^4$ and the correlation function $G^{(1)}$ for $K\!=\!3.3$ and for $\gamma\rightarrow\infty$ (left panels) and $\gamma\!=\!0$ (right panels) for different kick numbers ranging from $N_\mathrm{p}\!=\!0$ to 1501 as indicated. The black and blue dashed lines in (e) are fits with algebraic decay $\propto1/\sqrt{z}$ and exponential decay $\propto e^{-z/r_c}$, respectively. The red dashed line in (f) denotes a Lorentzian fit.
}
\label{suppfig:Num_G1&nk}
\end{figure}

\subsection{Localization for various trapping conditions}

Figure~\ref{suppfig:Num_trap} illustrates the evolution of the kinetic energy for two kick strengths and for three different trapping conditions, namely no trap, a flat-bottom trap as it is used throughout the manuscript, and a Gaussian trap with a harmonic trapping frequency of $14.7$ Hz as it is generated by the finite transversal extent of the lattice beams. Remarkably, the kinetic energy exhibits saturation at a later time for both $\gamma\rightarrow\infty$ and $\gamma\!=\!0$ under all trapping conditions, indicating the MBDL and DL phase, respectively. Note that the localized energy and onset time increase as the trap potential becomes steeper. Moreover, we find that the localized energy varies for different kick strengths, consistent with the experimental measurements shown in Fig.~\ref{fig:energy_entropy} in the main text. For all cases, when compared to the Gaussian trap, the flat-bottom trap leads to an earlier onset time of the localization and clearly decreases the localized energy. It thus helps us to observe the MBDL phase without notable heating and atom losses.


\subsection{Convergence of the energy with basis size}
To confirm the accuracy of our numerical results, we try four different basis sizes to calculate the evolution of the kinetic energy for $\gamma\!=\!0$ and $\gamma\!\rightarrow\!\infty$, as shown in Fig.~\ref{suppfig:Num_dimension}. The energy evolution is consistent with increasing basis size after $N_\mathrm{dim}\!=\!10000$. The localized energy is thus constant at a basis size beyond $N_\mathrm{dim}\!=\!10000$, see inset in Fig.~\ref{suppfig:Num_dimension}. Therefore, we choose $N_\mathrm{dim}=10000$ for all numerical simulations.

\subsection{MBDL for different atom numbers}
In our experiment, for the data presented in Fig.~\ref{fig:energy_entropy}, the total number of atoms is not constant. It drops by up to $25\%$ after one thousand kicks. In Fig.~\ref{suppfig:Num_particle_number}(a), we present the evolution of the energy per particle for $\gamma\rightarrow\infty$ for varying number of atoms $N$ ranging from $10$ to $33$. Since atoms cannot occupy the same eigenstate initially, the energy overall increases with increasing atom number at first, but it starts to stabilize at $N$ beyond $15$. The energy always saturates for all atom numbers in this range. As expected, the averaged localized energy exhibits a slight growth with increasing $N$ and is then followed by a fluctuation around $9E_\mathrm{r}$ (Fig.~\ref{suppfig:Num_particle_number}(b)). The increase of the localized energy with more atoms in the early stage can be attributed to the Pauli exclusion principle and the finite trap depth. These results indicate that MBDL is expected to remain robust in the thermodynamic limit $N\rightarrow\infty$.

\subsection{Momentum and $G^{(1)}$ evolution}
The calculated momentum distributions after different number of kicks for $\gamma\rightarrow\infty$ and $\gamma\!=\!0$ are shown in Fig.~\ref{suppfig:Num_G1&nk} (a) and (b). Remarkably, the profiles for both scenarios remain frozen after about $501$ kicks, and the momentum distributions for $\gamma\rightarrow\infty$ are much broader than those for $\gamma\!=\!0$. In Figs.~\ref{suppfig:Num_G1&nk}(c) and (d), we investigate the $n(k)k^4$ dynamics for both two scenarios to show Tan's contact~\cite{tan2008b,tan2008large} from theory side. As expected, the momentum distribution of a TG gas in equilibrium exhibits a power-law decay $n(k)=\mathcal{C}k^{-4}$ with $\mathcal{C}$ the Tan's constant, reflecting the local inter-atom interactions. As the kick number increases,  the plateau does not perfect due to the emergence of recoil peaks. Once MBDL has been occurred, the quantity $n(k)k^4$ fluctuates around a constant for momenta beyond $6k_\mathrm{L}$. We notice that this constant value is much higher than the one before kicking, which is caused by the fact that the kicks enhance the interactions between the atoms. In contrast, $n(k)k^4$ for $\gamma\!=\!0$ oscillates at high frequency over a wide range instead of being a constant.

We finally turn to the quantum correlations of the gases in the DL and MBDL phase. As we compare the one-body correlation function for the non-interacting and strongly interacting scenarios in the main text. In the ground state, the TG gas is expected to feature algebraic correlations $(G^{(1)}\propto1/\sqrt{z})$, reflecting the quasi-long-range order~\cite{cazalilla2011,Vuatelet2021effectivethermalization}. Thus, we numerically calculate $G^{(1)}$ for different kick numbers ranging from the ground state to the MBDL regime. As shown in Fig.~\ref{suppfig:Num_G1&nk}(e), the $G^{(1)}$ of an equilibrium TG gas is well fitted by an algebraic decay, but it gradually turns into an exponential decay $(G^{(1)}\propto e^{-z/r_c})$ after hundreds of kicks. In addition, the correlation length $r_\mathrm{c}$ is frozen after the kick number $N_\mathrm{p}\!=\!501$, which is consistent with the observed onset of MBDL. In contrast, the $G^{(1)}$ for $\gamma\!=\!0$ at different kick numbers are shown in Fig.~\ref{suppfig:Num_G1&nk}(f). Once DL has occurred, $G^{(1)}$ exhibits a Lorentzian decay indicating an exponential structure in momentum space. These calculations are consistent with our experimental results presented in the main text. This again highlights the many-body effect. \\

\end{twocolumngrid}


\begin{thebibliography}{53}
\bibitem [{Zen()}]{Zenodo}%
  \BibitemOpen
  \bibinfo {note} {{Data set is available from {Z}enodo at doi:
  10.5281/zenodo.10375982}}\BibitemShut {NoStop}%
\end{thebibliography}
\end{document}